\documentclass[aps,prx,twocolumn,showpacs,floatfix,superscriptaddress]{revtex4-1}

\usepackage[caption=false]{subfig}
\usepackage{verbatim}
\usepackage{graphicx}
\usepackage{dcolumn}
\usepackage{bm}
\usepackage{color}
\usepackage[colorlinks=true, allcolors=blue]{hyperref}
\usepackage{makeidx}
\usepackage{amsmath}
\usepackage{amssymb}
\usepackage{braket}
\usepackage{mathrsfs}
\usepackage{soul}
\makeindex

\renewcommand{\epsilon}{\varepsilon}
\newcommand{\up}{\uparrow}
\newcommand{\down}{\downarrow}

\begin{document}
\title{Levitons in superconducting point contacts}
\author{Matteo Acciai}
\affiliation{Dipartimento di Fisica, Universit\`a di Genova, Via Dodecaneso 33, 16146, Genova, Italy}
\affiliation{SPIN-CNR, Via Dodecaneso 33, 16146, Genova, Italy}
\affiliation{Aix Marseille Univ, Universit\'e de Toulon, CNRS, CPT, Marseille, France}
\author{Flavio Ronetti}
\affiliation{Dipartimento di Fisica, Universit\`a di Genova, Via Dodecaneso 33, 16146, Genova, Italy}
\affiliation{Aix Marseille Univ, Universit\'e de Toulon, CNRS, CPT, Marseille, France}
\author{Dario Ferraro}
\affiliation{Dipartimento di Fisica, Universit\`a di Genova, Via Dodecaneso 33, 16146, Genova, Italy}
\affiliation{SPIN-CNR, Via Dodecaneso 33, 16146, Genova, Italy}
\author{J\'er\^ome Rech}
\affiliation{Aix Marseille Univ, Universit\'e de Toulon, CNRS, CPT, Marseille, France}
\author{Thibaut Jonckheere}
\affiliation{Aix Marseille Univ, Universit\'e de Toulon, CNRS, CPT, Marseille, France}
\author{Maura Sassetti}
\affiliation{Dipartimento di Fisica, Universit\`a di Genova, Via Dodecaneso 33, 16146, Genova, Italy}
\affiliation{SPIN-CNR, Via Dodecaneso 33, 16146, Genova, Italy}
\author{Thierry Martin}
\affiliation{Aix Marseille Univ, Universit\'e de Toulon, CNRS, CPT, Marseille, France}
\date{\today}
\begin{abstract}
We investigate the transport properties of a superconducting quantum point contact in the presence of an arbitrary periodic drive. In particular, we calculate the dc current and noise in the tunnel limit, obtaining general expressions in terms of photoassisted probabilities. Interesting features can be observed when the frequency is comparable to the gap. Here, we show that quantized Lorentzian pulses minimize the excess noise, further strengthening the hierarchy among different periodic drives observed in the electron quantum optics domain. In this regime, the excess noise is directly connected to the overlap between electron and hole energy distributions driven out of equilibrium by the applied voltage. In the adiabatic limit, where the frequency of the drive is very small compared to the superconducting gap, we recover the conventional Shapiro-spikes physics in the supercurrent.
\end{abstract}

\maketitle
\section{Introduction}
Since the very first years following Josephson's prediction in 1962~\cite{josephson62}, electronic transport through coupled superconductors has been widely studied~\cite{josephson64review,waldram76review,likharev79review}. Later on, thanks to the advances in nanofabrication processes, it became possible to realize the so-called superconducting quantum point contacts~\cite{beenakker91,beenakker92,scheer97,goffman00,urbina12,glazman19heat} (SQPCs), i.e.\ systems where two superconducting electrodes are connected by a narrow constriction whose length is much smaller than the superconducting coherence length.
SQPCs are usually fabricated by relying on the break junction technique~\cite{moreland85,muller92,ruitenbeek96,vanRuitenbeek1997}, which paved the way to the realization of several experiments in this field \cite{ludoph00,cron01,scheer01,chauvin06,chauvin07} (see also Ref. \onlinecite{urbina12} for a broader overview on the subject). In addition, the implementation of a SQPC with split gate technology was very recently reported~\cite{thierschmann2018ncomm}.
In the mid 90s a unified theoretical approach describing normal metal-superconductor and superconductor-superconductor junctions under the effect of a constant voltage bias was developed~\cite{cuevas96prb}. In this context, multiple Andreev reflections~\cite{klapwijck82mar,blonder82mar} have been identified as the key ingredient to explain the subgap structure experimentally observed in the current-voltage characteristic. Several additional efforts have been put in the study of such junctions under the effect of microwave radiation, from early experiments by Shapiro~\cite{shapiro63} until much more recent research activity~\cite{cuevas02prl,stefanucci10,bergeret10,bergeret11,li18prb}, witnessing the interest in this topic.

On the other hand, a fast development of the so-called electron quantum optics (EQO)~\cite{grenier11,bocquillon12eqo,bocquillon14,bauerle2018review} occurred in the last decade. This very interesting research field aims at implementing the condensed matter counterpart of quantum optic setups. To achieve such a goal it is necessary to coherently generate and manipulate few-electron states. In this respect major advances are represented by the mesoscopic capacitor source~\cite{feve07,moskalets08meso,calzona16energypart} and quantized Lorentzian pulses~\cite{dubois2013levitonsNature,jullien14tomography}, recently implemented experimentally following earlier theoretical proposals~\cite{levitov96,levitov97}. In particular, predicted properties of the Lorentzian drive were confirmed by measuring the current noise produced when excitations generated by a periodic train of pulses are partitioned by a quantum point contact acting as a beamsplitter.
By relying on these tools and on the natural platform of quantum Hall edge states, several experiments have been performed~\cite{bocquillon13homscience,bocquillon2013,freulon15hom,tewari2016,marguerite16decoherence}, accompanied by an intense theoretical activity~\cite{degiovanni2009relaxation,degiovanni2010relaxation,jonckheere12hom,battista12,ferraro14decoherence,wahl14prl,dasenbrook15,moskalets15,ferraro15,ronetti16,moskalets16,moskalets17,ronetti17polarized,vannucci17heat,misiorny18,fleckenstein2018,ronetti18crystallization,cabart18,ferraro2018review,dashti19}. Among the most notable experimental achievements it is worth mentioning the implementation in condensed matter of the famous Hanbury-Brown and Twiss~\cite{hbt56} and Hong-Ou-Mandel~\cite{hom87} setups. All these studies show how current noise in the presence of an ac drive is an essential and well-established tool in EQO.

Electronic correlations associated with Coulomb interaction have been addressed in the context of EQO, mostly in the quantum Hall regime \cite{wahl14prl,rech16prl,acciai18}. It is therefore relevant to extend these concepts to superconducting devices, where correlations have a totally different nature.
In this paper we investigate a superconducting tunnel junction subject to an arbitrary periodic drive. In particular, we are interested in calculating the dc current and noise, for which we obtain general expressions in the framework of photoassisted transport~\cite{tien-gordon63,lesovik94,kouwenhoven94,pedersen98}. Indeed, while current has been widely studied in the literature, both in the presence of dc and ac drive~\cite{cuevas96prb,cuevas02prl}, noise is more often considered only in the presence of a dc bias~\cite{cuevas98prl,cuevas04fcs} and less attention has been dedicated to the more general case where a combined dc and ac drive is present. This is one of the main points we consider in this work. We ultimately have in mind to investigate the effects of superconducting correlations on Lorentzian voltage pulses, which play a major role in EQO.
Peculiar features of Levitons do emerge also in this case insofar as they minimize the excess noise due to quasiparticle transfers across the superconducting junction. These sharp differences between Levitons and other signals are best displayed when the driving frequency is {comparable to} the superconducting gap. 
In the opposite regime, where the superconducting gap is by far the dominant energy scale, we find for any drive a conventional Shapiro-spike structure in the supercurrent~\cite{shapiro63,waldram76review}, the main difference being in the height of the spikes which is related to a drive-dependent photoassisted amplitude.

The paper is organized as follows. Sec.\ \ref{sec:model} introduces the model for describing transport properties of the superconducting quantum point contact. We then present general expressions for the dc current and noise in Sec.\ \ref{sec:results}. Next, we discuss the peculiarities of Lorentzian pulses (Sec.\ \ref{sec:lor}), analyze the adiabatic limit (Sec.\ \ref{sec:adiabatic}) and present our conclusions in Sec.\ \ref{sec:conclusions}. Two Appendices are dedicated to technical details.
Throughout the {whole} paper we set $\hbar=1$.

\section{Model}\label{sec:model}
\begin{figure}[t]
	\begin{center}
		\includegraphics[width=\columnwidth]{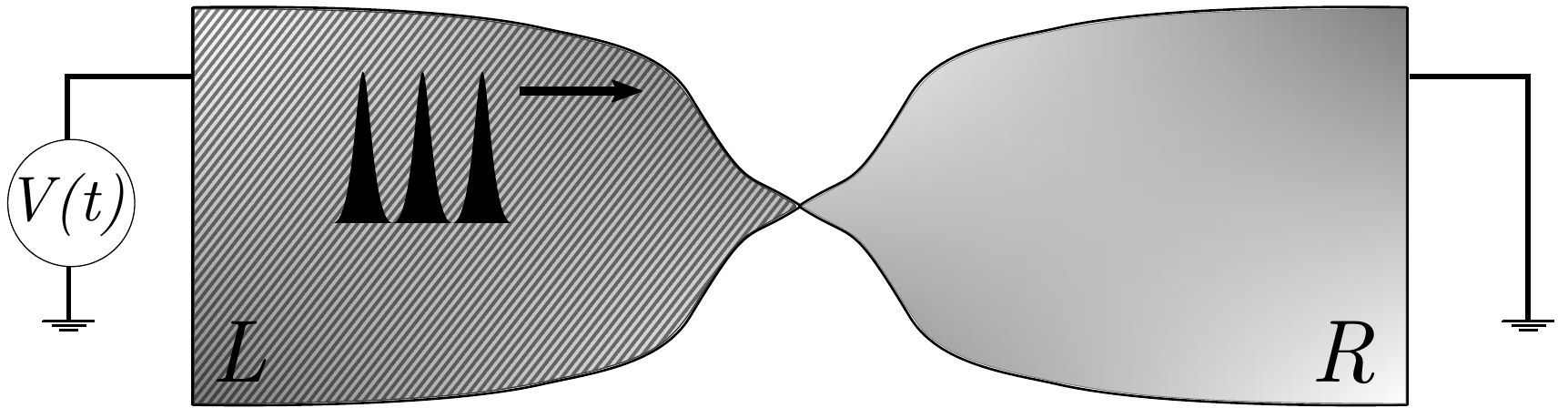}
	\end{center}
	\caption{Sketch of the considered setup. A narrow constriction between two superconducting electrodes implements a quantum point contact geometry. A time-dependent voltage $V(t)$ is applied to the left side of the junction {(shaded region)}, while the right electrode is grounded.}
	\label{fig:setup}
\end{figure}
In this paper we consider a driven SQPC~\cite{urbina12,beenakker92}, namely two superconducting electrodes connected by a narrow constriction whose length is much smaller than the superconducting coherence length. A periodic time-dependent voltage $V(t)=V_\text{dc}+V_\text{ac}(t)$ with angular frequency $\Omega=2\pi\mathcal{T}^{-1}$ is applied across the junction, as schematically depicted in Fig.\ \ref{fig:setup}. Here $V_\text{dc}$ is the dc contribution and $V_\text{ac}(t)$ the ac part having a vanishing average over one period $\mathcal{T}$. We adopt the model developed in Ref.\ \onlinecite{cuevas96prb}, according to which the essential features of our system can be described by considering a single quantum channel, with the following Hamiltonian:~\cite{rodero94prl,levy95prb,cuevas96prb}
\begin{equation}
H(t)=H_L+H_R+\lambda\sum_{\sigma=\up,\down}\left(e^{i\phi(t)}c_{L\sigma}^\dagger c_{R\sigma}+\text{H.c.}\right)\,.
\label{eq:H}
\end{equation}
Here, $H_L$ and $H_R$ are the BCS Hamiltonians of the uncoupled superconducting electrodes~\cite{bcs57} and the tunnel term accounts for electron transfers between them. We consider a symmetric junction, i.e.\ the {modulus of the} superconducting gap $\Delta$ is assumed to be the same in both right and left parts. Due to the presence of an external bias, hopping amplitudes are time dependent~\cite{rogovin74,barone} (see also App.\ \ref{app:distributions}) and characterized by the phase term {$\phi(t)=-\phi_0/2+e\int_{0}^{t}dt'\,V(t')$}, where $\phi_0$ is the bare superconducting phase difference between the electrodes and $e$ the electronic charge.

The average current across the junction is given by
\begin{equation}
I(t)=ie\lambda\sum_{\sigma=\up,\down}\left(e^{i\phi(t)}\Braket{c_{L\sigma}^\dagger (t)c_{R\sigma}(t)}-\text{H.c.}\right)\,,
\end{equation}
whereas the zero-frequency noise is defined as
\begin{equation}
S(t)=2\int_{-\infty}^{+\infty} dt' C(t+t',t)\,,
\end{equation}
with $C(t,t')=\Braket{I(t)I(t')}-\Braket{I(t)}\Braket{I(t')}$. Both current and noise can be expressed via nonequilibrium Keldysh Green's functions~\cite{keldysh64,rammer86,kamenev09} as a trace in Nambu space in the following way:~\cite{cuevas96prb,cuevas98prl}
\begin{equation}
I(t)=e\text{Tr}[\hat{\sigma}_3\hat{\mathcal{W}}(t)\hat{G}_{RL}^{+-}(t,t)-\hat{\sigma}_3\hat{G}_{LR}^{+-}(t,t)\hat{\mathcal{W}}^\dagger(t)]\,,\label{eq:i-start}
\end{equation}
\begin{align}
\begin{split}
C(t,t')&=2e^2\text{Tr}\left[\hat{\sigma}_3\hat{\mathcal{W}}(t)\hat{G}_{RR}^{-+}(t,t')\hat{\sigma}_3\hat{\mathcal{W}}^\dagger(t')\hat{G}_{LL}^{+-}(t',t)\right.\\
&\quad-\hat{\sigma}_3\hat{\mathcal{W}}(t)\hat{G}_{RL}^{-+}(t,t')\hat{\sigma}_3\hat{\mathcal{W}}(t')\hat{G}_{RL}^{+-}(t',t)\\
&\quad+\hat{\sigma}_3\hat{\mathcal{W}}^\dagger(t)\hat{G}_{LL}^{-+}(t,t')\hat{\sigma}_3\hat{\mathcal{W}}(t')\hat{G}_{RR}^{+-}(t',t)\\
&\quad\left.-\hat{\sigma}_3\hat{\mathcal{W}}^\dagger(t)\hat{G}_{LR}^{-+}(t,t')\hat{\sigma}_3\hat{\mathcal{W}}^\dagger(t')\hat{G}_{LR}^{+-}(t',t)\right]\,,
\end{split}
\label{eq:s-start}
\end{align}
where $\hat\sigma_3$ is the third Pauli matrix,
\begin{equation}
\hat{\mathcal{W}}(t)=
\begin{pmatrix}
\lambda\,e^{i\phi(t)} & 0\\
0 & -\lambda\,e^{-i\phi(t)}
\end{pmatrix}
\label{eq:w}
\end{equation}
and Green's functions are defined as ($i,j=R,L$)
\begin{equation}
\hat{G}_{i,j}^{+-}(t,t')=i
\begin{pmatrix}
\Braket{c_{j\up}^\dagger(t')c_{i\up}(t)} & \Braket{c_{j\down}(t')c_{i\up}(t)}\\
\Braket{c_{j\up}^\dagger(t')c_{i\down}^\dagger(t)} & \Braket{c_{j\down}(t')c_{i\down}^\dagger(t)}
\end{pmatrix}
\end{equation}
and $\hat{G}_{i,j}^{-+}(t,t')=[\hat{G}_{j,i}^{+-}(t,t')]^\dagger$. By treating the coupling term $\lambda$ in Eq.~\eqref{eq:H} as a perturbation, we obtain Green's functions from Dyson's equations involving unperturbed Green's functions $\hat{g}$ of the uncoupled electrodes (see App.\ \ref{app:is} for more details).
In the energy domain, the advanced and retarded components are~\cite{cuevas96prb}
\begin{equation}
\hat{g}^{a/r}(\omega)=\frac{1}{w\sqrt{\Delta^2-(\omega\mp i\epsilon)^2}}
\begin{pmatrix}
-\omega\mp i\epsilon & \Delta\\
\Delta & -\omega\mp i\epsilon
\end{pmatrix}\,,
\label{eq:greens}
\end{equation}
where $\epsilon=0^+$ and the energy scale $w\sim 1/\pi\rho(\epsilon_\text{F})$ is related to the normal density of states at the Fermi energy \cite{cuevas96prb}.
Other components of Green's functions are related to the above ones by $\hat{g}^{+-}(\omega)=2i\text{Im}[\hat{g}^a(\omega)]n_\text{F}(\omega)$ and $\hat{g}^{-+}(\omega)=-2i\text{Im}[\hat{g}^a(\omega)]n_\text{F}(-\omega)$, with $n_\text{F}(\omega)$ the Fermi function.

\section{Dc current and noise}\label{sec:results}
In this Section we present our results for the dc current and noise. These quantities are defined as a time average of $I(t)$ and $S(t)$ over a {measurement time $\mathscr{T}$ (much longer than all the other time scales in the system)}, i.e.\
$I=\mathscr{T}^{-1}\int_{-\mathscr{T}/2}^{\mathscr{T}/2}dt\,I(t)$ and likewise for the noise. We consider the tunnel regime where the transmission of the junction is very small, so that current and noise can be evaluated to lowest order in the tunneling amplitude $\lambda$. The result can be expressed as:
\begin{subequations}
\begin{align}
I&=I_0+\chi_{2q}(I_1+I_J)\,,\label{eq:res-i}\\
S&=S_0+\chi_{2q}S_1\,,\label{eq:res-s}
\end{align}
\label{eq:res-is}
\end{subequations}
where $\chi_x=1$ if $x\in\mathbb{Z}$ and zero otherwise, while $q\Omega=eV_{\text{dc}}$, with $V_\text{dc}=\mathcal{T}^{-1}\int_0^{\mathcal{T}}dt\,V(t)$ the dc component of the drive (recall that $\mathcal{T}$ is the period).

All contributions can be expressed in terms of the photoassisted amplitudes~\cite{dubois13prb,ferraro18squeezing}
{\begin{equation}
p_{\ell}(\alpha) = \int_{-\mathcal{T}/2}^{\mathcal{T}/2} \frac{dt}{\mathcal{T}} e^{2 i \pi  \ell \frac{t}{\mathcal{T}}} e^{- 2 i\pi  \alpha  \Phi(t)}, 
\label{eq:photo}
\end{equation}
with 
\begin{equation}
\Phi (t) = \int_{0}^t \frac{dt'}{\mathcal{T}}\bar{V}_{\text{ac}}(t')
\end{equation}
where $\bar{V}_{\text{ac}}(t)$ is the ac part of $V(t)$ with unitary and dimensionless amplitude.}
Here, by analogy with $q$, we introduced a parameter $\alpha=eV_\text{ac}^0/\Omega$, where $V_\text{ac}^0$ is the characteristic amplitude of the ac component of the drive. For instance, in the case of a harmonic drive, {$V(t)=V_\text{dc}+V_\text{ac}^0\cos(\Omega t)$}. Coefficients in Eq.~\ \eqref{eq:photo} represent the probability amplitude for an electron to emit $(\ell<0)$ or absorb $(\ell>0)$ $|\ell|$ photons of energy $\Omega$ as a consequence of the ac drive \cite{dubois13prb}. At low but finite temperature, terms in Eq.~\eqref{eq:res-is} can be expressed as a single integral over energies (see App.\ \ref{app:is}), while analytic results are found at zero temperature. In this case, terms appearing in the current Eq.~\eqref{eq:res-i} are
\begin{align}
I_0&=\frac{4e\lambda^2}{\pi w^2}\sum_{\ell\in\mathbb{Z}}|p_\ell|^2\Theta(1-|\Delta_\ell|)\,\Omega_\ell\,\mathcal{J}(\Delta_\ell)\label{eq:i0}\\
\frac{I_1}{\Delta}&=-\frac{4e\lambda^2}{\pi w^2}\sum_{\ell\in\mathbb{Z}}\text{Re}[e^{i\phi_0}p_\ell p_{-\ell-2q}]\Theta(1-|\Delta_\ell|)\Delta_\ell K(\tilde{\Delta}_\ell)\label{eq:i1}\\
\begin{split}
\frac{I_J}{\Delta}&=\frac{4e\lambda^2}{\pi w^2}\sum_{\ell\in\mathbb{Z}}\text{Im}[e^{i\phi_0}p_\ell p_{-\ell-2q}]|\Delta_\ell|\\
\times&\left[\Theta(1-|\Delta_\ell|) K(\Delta_\ell)-i\Theta(|\Delta_\ell|-1)F\left(\varphi_\ell,\tilde{\Delta}_\ell\right)\right]
\end{split}\label{eq:ij}
\end{align}
where $\Theta(x)$ is the Heaviside step function, $\Omega_\ell=(\ell+q)\Omega$, $\Delta_\ell=2\Delta/\Omega_\ell$, $\tilde{\Delta}_\ell=\sqrt{1-\Delta_\ell^2}$, $\varphi_\ell=\sin^{-1}(1/\tilde{\Delta}_\ell)$, $F(\varphi,x)$ the incomplete elliptic integral of the first kind and $\mathcal{J}(x)=E(\sqrt{1-x^2})-x^2K(\sqrt{1-x^2})/2$, with $K(x)$ and $E(x)$ the complete elliptic integrals of the first and second kind, respectively \cite{gradshteyn2014}.
Expressions for noise contributions in Eq.~\eqref{eq:res-s} are quite similar:
\begin{align}
S_0&=\frac{8e^2\lambda^2}{\pi w^2}\sum_{\ell\in\mathbb{Z}}|p_\ell|^2\Theta(1-|\Delta_\ell|)\,|\Omega_\ell|\,\mathcal{J}(\Delta_\ell)\,,\label{eq:s0}\\
\frac{S_1}{\Delta}&=-\frac{8e^2\lambda^2}{\pi w^2}\sum_{\ell\in\mathbb{Z}}\text{Re}[e^{i\phi_0}p_\ell p_{-\ell-2q}]\Theta(1-|\Delta_\ell|)|\Delta_\ell| K(\tilde{\Delta}_\ell)\,.\label{eq:s1}
\end{align}
\begin{figure}[t]
	\begin{center}
		\includegraphics[width=\columnwidth]{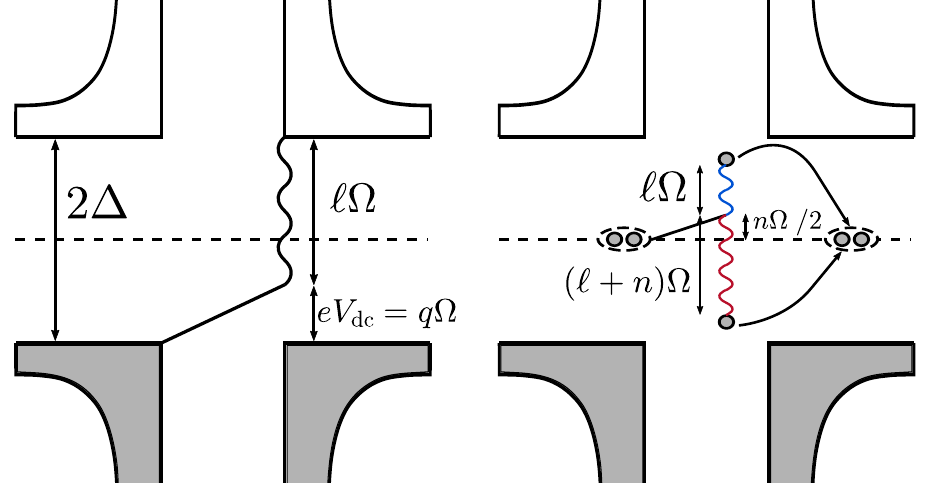}
	\end{center}
	\caption{{Sketch of typical processes involved in the dc current. Left panel: out-of gap process contributing to $I_0$. A quasiparticle gains an energy $q\Omega=eV_\text{dc}$ (straight line) from the dc part of the drive and absorbs $\ell$ photons (wiggly line) to overcome the energy gap, thanks to the additional energy contribution $\ell\Omega$. This process is weighted by the probability $|p_\ell|^2$, appearing in the expression for $I_0$. Right panel: sub-gap process contributing to $I_J$. This process globally results in a transfer of a Cooper pair. Both electrons gain from the dc part of the drive an energy $q\Omega$, with $q=n/2$, $n\in\mathbb{N}$. Then the process is an interference between one electron absorbing $\ell$ photons (with amplitude $p_\ell$, $\ell>0$) and the other emitting $\ell+n$ photons (with amplitude $p_{-\ell-2q}=p_{-\ell-n}$). Finally the two electrons recombine to form a Cooper pair.}}
	\label{fig:processes}
\end{figure}
Note that all expressions above apply for arbitrary periodic drives, as the nature of the drive is solely encoded in the $p_\ell$ coefficients.
Let us now comment on results in Eq.~\eqref{eq:res-is} and their explicit expressions given below. Both the current and the noise contain a {continuous} contribution as a function of $q$ ($I_0$ and $S_0$) and terms appearing only at discrete values of the dc voltage, namely when $2q$ is integer. The latter are Shapiro step contributions~\cite{shapiro63,waldram76review} and are due to the interplay of the ac Josephson effect and the frequency $\Omega$ of the external drive, that together give rise to a dc contribution (inverse ac Josephson effect). The external bias appears in all terms via the combination $\Omega_\ell=(\ell+q)\Omega$, a typical signature of photoassisted transport.
\begin{figure}[t]
	\begin{center}
		\includegraphics[width=\columnwidth]{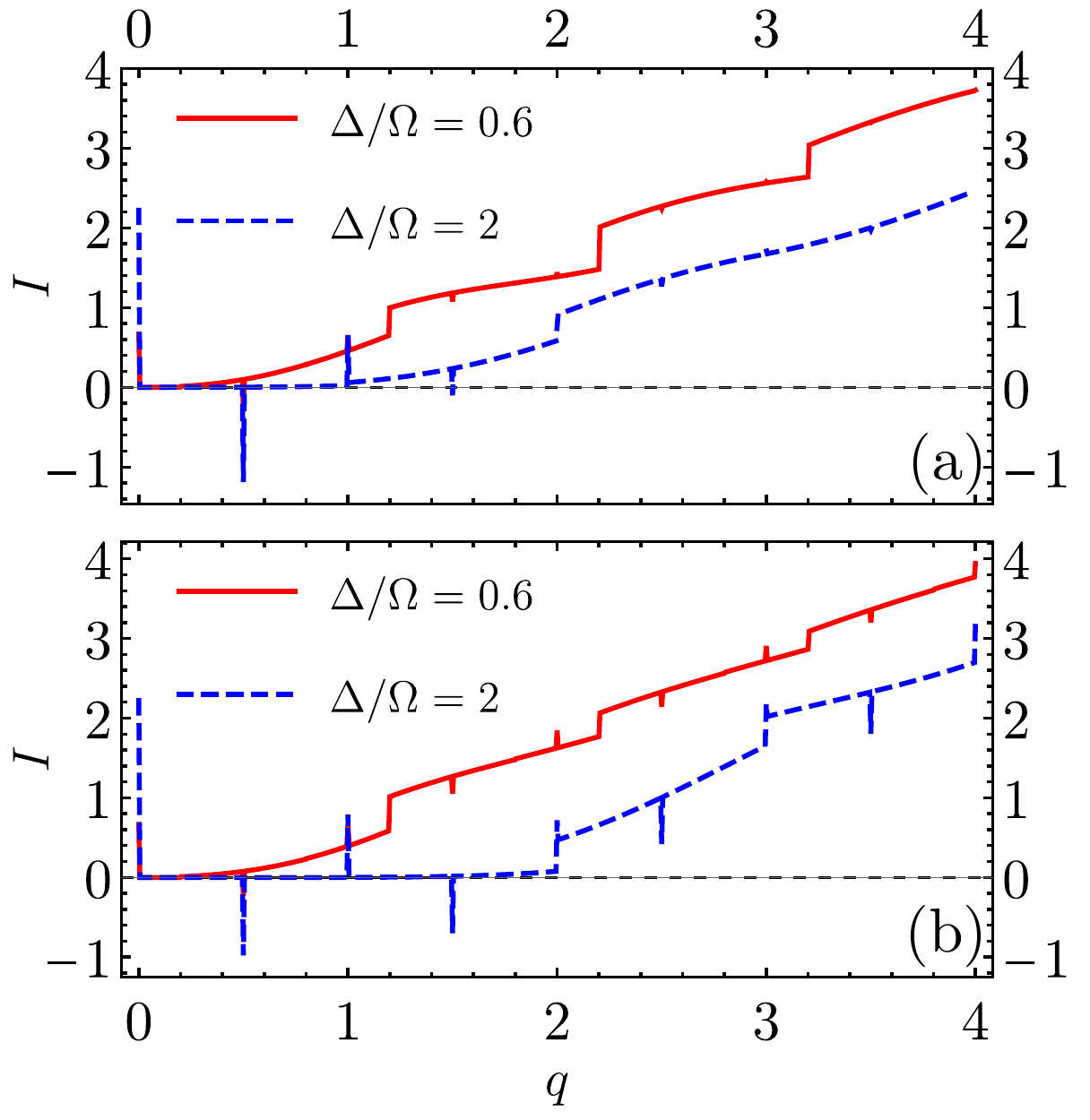}
	\end{center}
	\caption{Total current $I$ as a function of $q$, in units of $eT\Omega/\pi$ and for two values of $\Delta/\Omega$, as indicated in the plots. $T=4\lambda^2/w^2$ is the transmission of the junction. (a): the case of a Lorentzian drive with $\eta=0.1$ [see Eq.\ \eqref{eq:lor}]. (b): the case of a sine drive $V(t)=V_\text{dc}[1-\cos(\Omega t)]$. In both panels we set $\phi_0=\pi/4$.}
	\label{fig:current}
\end{figure}

$I_0$ represents the current due to quasiparticle transfers across the junction; it involves only out-of-gap processes (due to the $\Theta$ function enforcing the ``effective voltage'' $\Omega_\ell$ to be greater than $2\Delta$) and is independent of the superconducting phase difference $\phi_0$. {A typical process contributing to $I_0$ is depicted in Fig.\ \ref{fig:processes} (left panel).} It is easy to see that, in the metallic limit $\Delta=0$, $I_0$ is the only surviving contribution to the current and reduces to the well known result $I_0=T(2e^2/h)V_{\text{dc}}$~\cite{dubois13prb}, where $T=4\lambda^2/w^2$ is the transmission of the junction in the tunnel limit~\cite{cuevas96prb} and $2e^2/h$ is the conductance of a spinful quantum channel. Concerning the phase-dependent terms, $I_J$ is the only contribution involving also sub-gap processes [second $\Theta$ function in Eq.~\eqref{eq:ij}] and is a generalization of the dc Josephson current in the presence of an arbitrary periodic drive. {It involves a transfer of Cooper pairs across the junction. From the dependence $p_\ell p_{-\ell-2q}$ (see Eq.\ \eqref{eq:ij}), we can interpret each transfer as an interference between processes where an electron absorbs $\ell$ photons, with amplitude $p_\ell$ and another one emits $(\ell+n)$ photons, with amplitude $p_{-\ell-2q}$ and $2q=n$, which is the condition enforced by the factor $\chi_{2q}$ in Eq.\ \eqref{eq:res-is}. Since both electrons also gain an energy $q\Omega=n\Omega/2$ from the dc part of the voltage, we then see that the final energies of the two electrons are equal and opposite, so that they recombine into a Cooper pair. This kind of process is also sketched in Fig.\ \ref{fig:processes} (right panel).} In the limit of a purely dc bias, which is obtained by replacing $p_\ell=\delta_{\ell,0}$, $I_J$ reduces to $I_J=\delta_{q,0}T\frac{e\Delta}{2}\sin(\phi_0)$ and we recover the dc Josephson effect, with supercurrent flowing at zero bias~\cite{josephson62,barone}. Of course, $I_J$ is the only surviving contribution if no drive is applied to the system. The remaining term, $I_1$, {has the same origin as the contribution proportional to $\cos\phi_0$ in the ac Josephson effect and} can be interpreted as describing quasiparticle processes involving a superimposed pair transfer~\cite{langenberg74,barone}.

In Fig.\ \ref{fig:current} we show some examples of how the total current $I$ behaves as a function of $q$. We chose a Lorentzian and a sine drive, which will be discussed in detail in Sec.\ \ref{sec:lor} in relation to the excess noise. From the plots in Fig.\ \ref{fig:current} we clearly observe the continuous contribution $I_0$, characterized by some discontinuities due to the $\Theta$ functions in the sum in Eq.\ \eqref{eq:i0}. On top of that, Shapiro spikes at half-integer values of $q$ appear. They come almost completely from $I_J$, since $I_1$ is found to be negligible for a wide range of parameters.

Finally, concerning the noise, $S_0$ and $S_1$ are the counterparts to $I_0$ and $I_1$, respectively, and are {generated by the same processes contributing to $I_0$ and $I_1$. In particular, $S_0$ is} associated with the partitioning of quasiparticles excited above the gap by the driving voltage. There is however no term in the noise associated with sub-gap processes appearing in $I_J$, which are therefore noiseless~\cite{rogovin74,barone}. In the following we analyze the above general results in two different regimes.

\section{Excess noise and Lorentzian drive}\label{sec:lor}
Among all possible periodic drives, Lorentzian pulses play a special role since they are known to generate minimal excitations in conventional ballistic conductors~\cite{levitov96,levitov97,keeling06} and also in strongly correlated states such as the fractional quantum Hall effect~\cite{rech16prl}. For this reason they have been widely studied in the framework of EQO~\cite{grenier11,dubois2013levitonsNature,dubois13prb,vannucci17heat,ronetti18crystallization,acciai18,glattli18,glattli17,Dolcini2018}.
It is then natural to ask whether some of these signatures survive in the superconducting system we are considering in this paper. In what follows we first introduce the definition of excess noise for a generic drive and subsequently show how integer Levitons still lead to its minimization while other drives do not. In this Section we consider the ac and dc amplitudes of the drive to be equal, namely $\alpha=q$.

For a generic drive, the excess noise can be defined in the following way~\cite{dubois13prb,rech16prl}:
\begin{equation}
\Delta S=S-2eI\,.
\label{eq:excess-noise-def}
\end{equation}
It represents the deviation of the noise from its Poissonian limit~\cite{blanter00}. The above definition involves the total current and noise and can be decomposed as $\Delta S=\Delta S_0+\Delta S_1-2eI_J$, with $\Delta S_{0,1}=S_{0,1}-2eI_{0,1}$.
In particular, $\Delta S_0$ only refers to quasiparticle terms and will constitute the main focus of our discussion. As a matter of fact, $S_1$, $I_1$ and $I_J$ are defined only for half-integer values of $q$ and depend on the superconducting phase difference $\phi_0$. Therefore, in a setup where $\phi_0$ is not fixed, it is in principle possible to isolate $S_0$ and $I_0$. Indeed, $I_1$ and $S_1$ will vary as $\cos(\phi_0)$, while $I_J$ as $\sin(\phi_0)$ and then these contributions can be subtracted by averaging over different measurements.
For these reason we focus our attention on $\Delta S_0$.
From Eq.~\eqref{eq:i0} and Eq.~\eqref{eq:s0} we immediately find:
\begin{equation}
\Delta S_0=\frac{16e^2\lambda^2}{\pi w^2}\sum_{\ell <-q}|p_\ell|^2\Theta(1-|\Delta_\ell|)\Omega|\ell+q|\mathcal{J}(\Delta_\ell)\,.
\label{eq:delta-s0}
\end{equation}
\begin{figure*}
	\begin{center}
		\includegraphics[width=\textwidth]{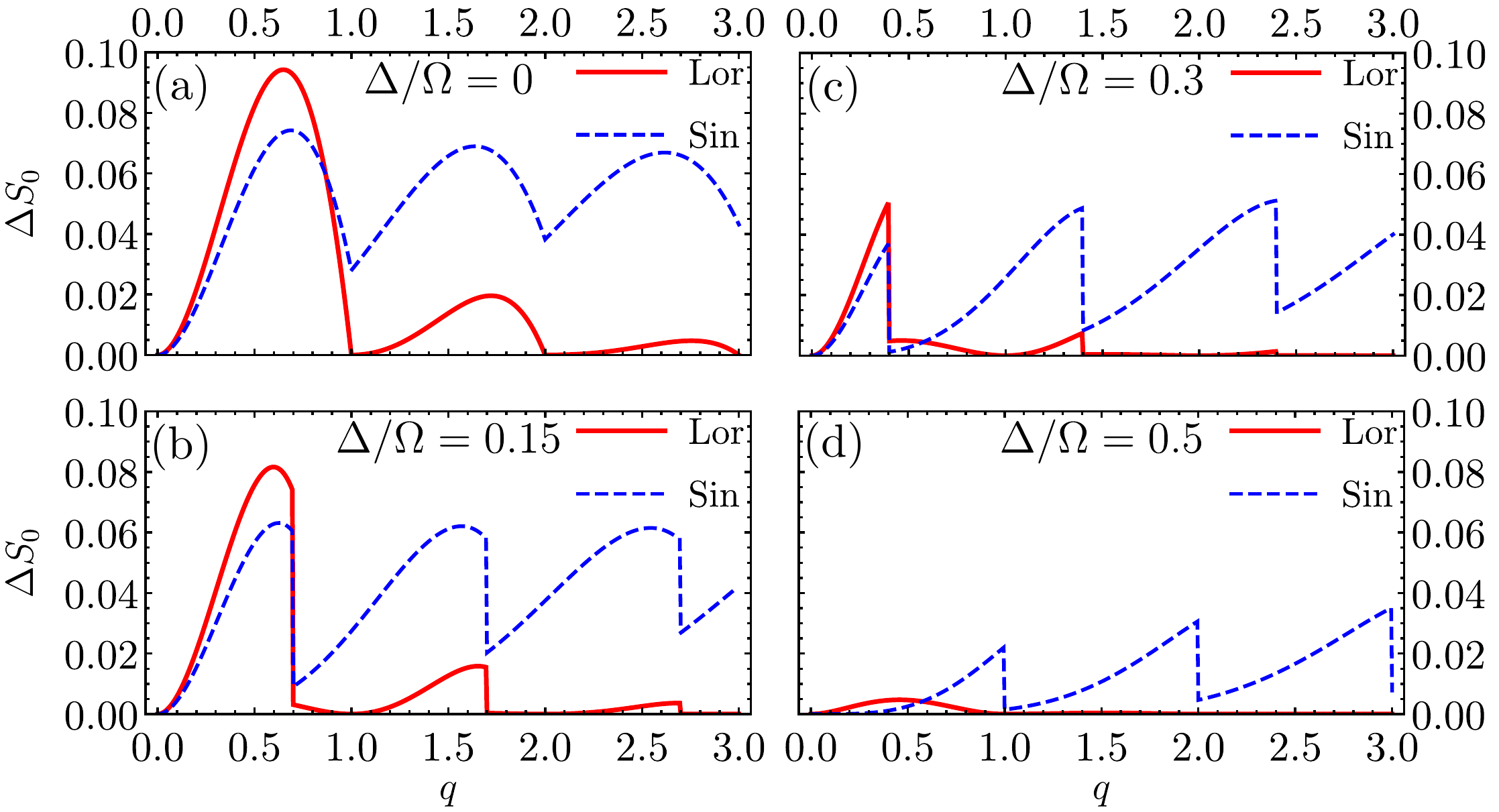}
		\caption{Excess noise $\Delta S_0$ for different values of $\Delta/\Omega$ as a function of $q$, in units of $2e^2T\Omega/\pi$. The width of Lorentzian pulses is $\eta=0.1$. Full red curves refer to Lorenzian pulses, dashed blue ones to a sine drive $V(t)=V_\text{dc}[1-\cos(\Omega t)]$, with $q\Omega=eV_\text{dc}$.}
		\label{fig:excess}
	\end{center}
\end{figure*}

Before moving to the discussion of Lorentzian pulses, we now highlight a deeper connection between the excess noise and single-electron properties. Very generally, by starting from Eq.~\eqref{eq:s-start} and using Dyson's equations \eqref{eq:dys1} and \eqref{eq:dys2}, one can show that the excess noise $\Delta S_0$ can be written in terms of Green's function as
\begin{equation}
\Delta S_0=\frac{4e^2\lambda^2}{\pi}\int d\omega\, g_0^{+-}(\omega)\sum_{\ell\in\mathbb{Z}}|p_\ell|^2g_0^{-+}(\omega-\Omega_\ell)\,.
\label{eq:excess-noise-green}
\end{equation}
Here, the subscript $0$ in Green's functions denotes the term proportional to the identity matrix $\hat{\sigma}_0$ in Nambu space. Recall also that $\Omega_\ell=(\ell+q)\Omega$. This formula has the typical structure of the Tien-Gordon effect~\cite{tien-gordon63} and involves an overlap between two Green's functions: $g_0^{+-}(\omega)$ at equilibrium and $g_0^{-+}(\omega)$, shifted by the dc bias $q\Omega$ as well as all energies $\ell\Omega$ corresponding to photoassisted processes and weighted by the probability $|p_\ell|^2$.

It is possible to link Eq.\ \eqref{eq:excess-noise-green} to electron energy distributions which are usually employed in the context of EQO~\cite{grenier11,ferraro13}. In particular, here we refer to nonequilibrium energy distribution of the left {side} of the SQPC. 
We refer to Appendix \ref{app:distributions} for the details and here we simply state the result:
\begin{equation}
\Delta S_0\propto\int d\omega\,f_\text{eq}^{(e)}(\omega)f^{(h)}(-\omega)\,.
\label{eq:excess-noise-f}
\end{equation}
Essentially, $g_0^{+-}(\omega)$ gives the electron energy distribution at equilibrium $f_{\text{eq}}^{(e)}(\omega)$, while the sum containing $g_0^{-+}(\omega-\Omega_\ell)$ represents the hole energy distribution $f^{(h)}(-\omega)$ in the presence of the drive. Explicit expressions at zero temperature are
\begin{equation}
\begin{split}
f^{(e)}_\text{eq}(\omega)&=\frac{-2\omega}{w\sqrt{\omega^2-\Delta^2}}\Theta(-\omega-\Delta)\,,\\
f^{(h)}(\omega)&=\sum_{\ell\in\mathbb{Z}}\frac{|p_\ell|^2}{w}\frac{-2(\omega+\Omega_\ell)}{\sqrt{(-\omega-\Omega_\ell)^2-\Delta^2}}\Theta(-\omega-\Omega_\ell-\Delta)\,.
\end{split}
\label{eq:energy-distr}
\end{equation}

As a final remark, we notice that a similar procedure can be followed for $\Delta S_1$. Indeed, despite this term being negligible in our discussion, it can be shown that (assuming real $p_\ell$)
	\begin{equation}
	\Delta S_1\propto\cos\phi_0\int d\omega\, g_1^{+-}(\omega)\sum_{\ell\in\mathbb{Z}}p_\ell p_{-\ell-2q}g_1^{-+}(\omega-\Omega_\ell)\,,
	\end{equation}
	where $g_1$ is the off-diagonal component of the Green's function in Nambu space. The above expression can be obtained starting from anomalous correlators of the form $\Braket{c_{L\down}(t')c_{L\up}(t)}$, by analogy with what is done in Appendix \ref{app:distributions}.
Let us now discuss in detail the relevant case of a Lorentzian drive. A train of Lorentzian-shaped pulses has the form
\begin{equation}
V(t)=\frac{V_\text{dc}}{\pi}\sum_{k\in\mathbb{Z}}\frac{\eta}{\eta^2+(t/\mathcal{T}-k)^2}\,,
\label{eq:lor}
\end{equation}
where $\eta$ is the ratio between the width of a pulse and the period $\mathcal{T}$ of the drive. Its photoassisted coefficients $p_\ell$ have been given in different references (see for instance Refs.\ \onlinecite{dubois13prb,rech16prl}) and have the peculiar property that they vanish for $\ell<-q$ in the case of quantized pulses, i.e.\ for integer values of $q$. This has the consequence that $I_1$ and $S_1$ are zero for integer Levitons. Indeed, the combination of photoassisted coefficients appearing in Eq.~\eqref{eq:i1} and Eq.~\eqref{eq:s1} becomes in this case $p_\ell p_{-\ell-2q}=\chi_q\delta_{l,-q}p_{-q}^2$, enforcing $\ell=-q$. {Therefore, $I_1=S_1=0$ due to the action of the $\Theta$ functions}. This means that, unlike any other drive, the noise for quantized Lorentzian pulses is independent of the bare superconducting phase difference $\phi_0$. Moreover, another interesting property is that the $I_J$ contribution reduces to
\begin{equation}
I_J=T\frac{e\Delta}{2}p^2_{-q}\sin(\phi_0)
\label{eq:jos-lev}
\end{equation}
for integer Levitons. This is a very simple Josephson-like relation, where supercurrent peaks occurring at integer $q$ are weighted by the photoassisted amplitude $p_{-q}^2$.

Concerning the behavior of the excess noise, Eq.\ \eqref{eq:delta-s0} shows that it vanishes for Levitons with integer charge, by analogy with what was observed in the free-electron case~\cite{dubois13prb,dubois2013levitonsNature}. This is a direct consequence of the properties of their $p_\ell$ coefficients. In Fig.\ \ref{fig:excess} we plot the excess noise $\Delta S_0$ for different values of the ratio $\Delta/\Omega$, comparing Lorentzian and {cosine} drives. In the metallic limit $\Delta=0$ [Fig.\ \ref{fig:excess}(a)] we recover known behaviors~\cite{dubois13prb,dubois2013levitonsNature}, while at finite gap we observe the appearence of sharp discontinuities [Figs.\ \ref{fig:excess}(b)--(d)] which are due to the BCS density of states, as we will argue in the following. Still, we clearly observe that quantized Lorentzian pulses minimize the excess noise, in contrast to the harmonic voltage. By increasing the ratio $\Delta/\Omega$, we observe a progressive overall suppression of the signal for both drives. This can be understood by noticing that, in the adiabatic limit $\Delta\gg\Omega,\,eV_\text{dc}$, no contribution other than $I_J$ can survive, since no transport across the gap is possible anymore and $I_J$ is the only term involving also sub-gap processes (see Sec.\ \ref{sec:adiabatic} for a more thorough discussion). For this reason, even though only quantized Levitons minimize the excess noise (strictly speaking), the major differences between integer Lorentzian pulses and any other drive are best appreciated if the ratio $\Delta/\Omega$ is at most of the order of unity. We comment about this constraint in Sec.\ \ref{sec:conclusions}. By increasing $\Delta/\Omega$, we progressively enter the adiabatic regime and the transport properties of the junction become qualitatively similar for any drive, as we will discuss in the following Section.

\begin{figure}[t]
	\begin{center}
		\includegraphics[width=\columnwidth]{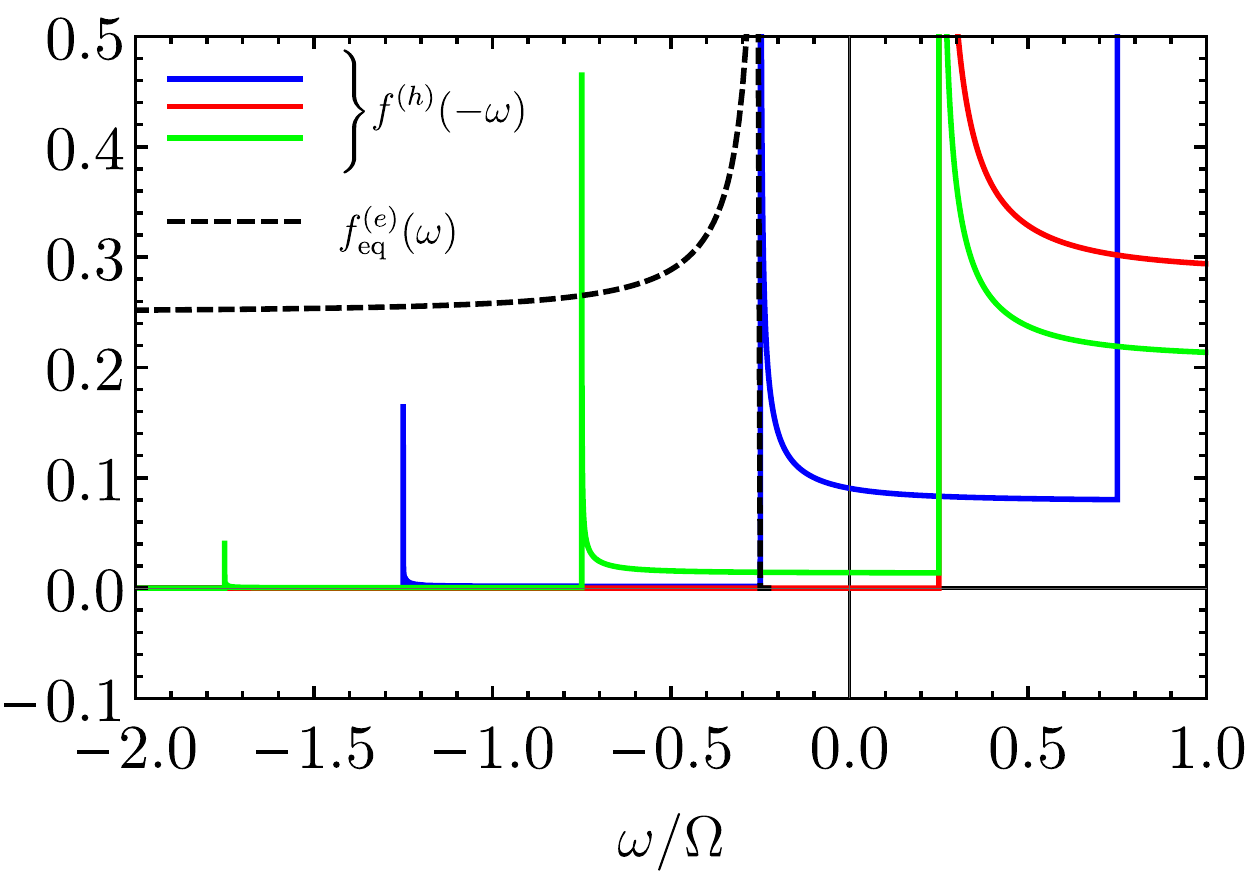}
		\caption{Overlap between equilibrium distribution $f_\text{eq}^{(e)}(\omega)$ (black dashed curve) and out-of-equilibrium distribution $f^{(h)}(-\omega)$ (both in units of $2/w$) for $\Delta/\Omega=0.25$ and: Lorentzian drive at $q=1$ (red curve), Lorentzian drive at $q=0.5$ (blue curve) and sine drive at $q=1$ (green curve). The width of Lorentzian pulses is $\eta=0.1$. {Notice that the} equilibrium distribution has been reduced by a factor 4 to better appreciate the contributions from $f^{(h)}(-\omega)$, which are quite small in the region $\omega<-\Delta$.}
		\label{fig:overlap}
	\end{center}
\end{figure}
Finally, we illustrate the behavior of distribution functions in Eq.\ \eqref{eq:energy-distr}, which are related to the excess noise by Eq.\ \eqref{eq:excess-noise-f}.
Fig.\ \ref{fig:overlap} shows the overlap of these distributions for some values of $q$ and a fixed $\Delta/\Omega$. It is always zero for quantized Levitons because in this case $p_\ell=0$ for $\ell<-q$. This means that $f^{(h)}(-\omega)$ is nonzero only for $\omega>\Delta+\ell\Omega$, with $\ell\ge 0$ and the overlap vanishes because $f^{(e)}_\text{eq}(\omega)$ is nonzero for $\omega<-\Delta$. This is no longer the case for non-quantized Lorentzians or any other drive, for which $f^{(h)}(-\omega)$ is nonvanishing also in the region $\omega<-\Delta$.
The structure of functions in Eq.~\eqref{eq:energy-distr} also allows us to understand the discontinuities observed in Fig.\ \ref{fig:excess}. Indeed, both $f^{(e)}_\text{eq}$ and $f^{(h)}$ show signatures of the square root singularity of the BCS density of states. The singularity of the equilibrium distribution is at $\omega=-\Delta$, while those of $f^{(h)}$ {depend} on the values of $\ell$ and $q$. When a singularity of $f^{(h)}(-\omega)$ enters/leaves the region $\omega<-\Delta$, an abrupt increase/decrease of the overlap between the two distribution occurs. At a given $\ell$, this happens when $q=-\ell-2\Delta/\Omega$, which are precisely the values where discontinuities in $\Delta S_0$ are observed (see Fig.\ \ref{fig:excess}).

\section{Adiabatic limit}\label{sec:adiabatic}
Let us now analyze the situation where the superconducting gap is the most relevant energy scale in the problem. This, in particular, means that both the excitation frequency $\Omega$ and $eV_\text{dc}$ have to be much smaller than the gap $\Delta$. In this limit all contributions to the current and noise but $I_J$ are {progressively} suppressed. Mathematically, this is because the bigger the gap, the higher the value that the index $\ell$ has to assume to prevent $\Theta$ functions from vanishing. Although $\ell$ can assume any value in principle, in practice contributions at high $\ell$ are {strongly suppressed due to the} $p_\ell$ coefficients. More physically and intuitively, this means that when $\Delta$ is by far the biggest energy scale, the drive cannot provide {enough energy to the system} for out-of-gap processes to be possible, even with the photoassisted tunneling mechanism. Therefore the relevant quantity in the adiabatic regime is the part of $I_J$ involving sub-gap processes. Thanks to the limit $\Delta\gg\Omega,\,eV_\text{dc}$, Eq.\ \eqref{eq:ij} for $I_J$ considerably simplifies and becomes
\begin{equation}
I_J=T\,\frac{e\Delta}{2}\sum_{\ell\in\mathbb{Z}}\text{Im}[p_\ell p_{-\ell-2q}e^{i\phi_0}]=T\frac{e\Delta}{2}p_{-2q}(2\alpha)\sin(\phi_0)\,,
\label{eq:ij-shapiro}
\end{equation}
where we assumed, without loss of generality, that photoassisted coefficients are real and we used the general property $\sum_{\ell}p_\ell(\alpha)p_{-\ell+x}(\alpha)=p_{x}(2\alpha)$.
This result has the same structure of Eq.\ \eqref{eq:jos-lev}, to which it reduces in the case of a Lorentzian drive, since $p_{-2q}(2q)=p_{-q}^2(q)$ for integer $q$. We emphasize, though, that in the case of integer Levitons Eq.\ \eqref{eq:jos-lev} holds for any value of the ratio $\Omega/\Delta$, without any restriction.
Eq.\ \eqref{eq:ij-shapiro} describes a series of supercurrent spikes appearing whenever $2q$ is integer, whose amplitude is determined by the photoassisted coefficient $p_{-2q}(2\alpha)$ (recall that $\alpha$ is related to the ac amplitude of the drive). The condition $2q\in\mathbb{Z}$ means that the dc amplitude of the drive has to satisfy $V_{\text{dc}}=k\Omega/2e$, with integer $k$. The appearence of Shapiro spikes in the $I-V$ characteristic in the presence of a harmonic drive is a well-known result and is due to the inverse ac Josephson effect~\cite{waldram76review}. Here, we recover the same kind of effect, but in the presence of an arbitrary periodic drive. The photoassisted coefficient $p_{-2q}(2\alpha)$ replaces and generalizes the usual Bessel function $(-1)^{k}J_k(2eV_\text{ac}^0/\Omega)$ that is found for a harmonic drive~\cite{waldram76review}, {$V(t)=V_\text{dc}+V_\text{ac}^0\cos(\Omega t)$} (with $k=2q$ an integer number).

Finally, we also notice that the relation in Eq.~\eqref{eq:ij-shapiro} could be used as a tool to operate a ``spectroscopy'' of photoassisted absorption and emission probabilities by varying independently $\alpha$ and $q$, in the same spirit of what has been proposed in Ref.\ \cite{dubois13prb}. It is indeed possible to vary the ac amplitude of the drive (and hence $\alpha$) in correspondence of the fixed dc amplitudes where Shapiro spikes occur, thus recovering $p_\ell$ coefficients from the amplitude of the spike.

\section{Conclusions}\label{sec:conclusions}
In this work, we have considered transport properties of a superconducting quantum point contact in the tunnel regime, in the presence of an arbitrary periodic drive. In particular, we calculated the dc current across the junction and the zero-frequency noise at lowest order in the tunneling amplitude, by relying on a nonequilibrium Keldysh Green's functions approach, and obtained general expressions in terms of photoassisted amplitudes.

When the angular frequency of the drive $\Omega$ is comparable to the superconducting gap $\Delta$, sharp differences between quantized Lorentzian pulses and every other signal occur. Indeed, the former drive is the only one for which the excess noise associated with quasiparticle processes vanishes. Remarkably enough, this well known property of ballistic metallic systems still persist when entering the superconducting regime. Moreover, the total noise becomes independent of the bare superconducting phase difference $\phi_0$.
This work therefore contributes to the characterization of single quasiparticle transfer between two superconductors, in the same spirit of what was previously achieved in EQO scenarios in the ballistic regime for single electron excitations.

From the experimental point of view, the constraint $\Delta/\Omega\lesssim 1$ is quite challenging but not unreachable. In SQPCs realized with the break junction technique, the typical regime is more towards the opposite case~\cite{chauvin06} (with the gap in the range of hundreds of $\mu$eV and $\nu=\Omega/2\pi$ in the range of a few tens of GHz). However, some recent experiments~\cite{thierschmann2018ncomm} are extremely promising to explore the $\Delta/\Omega\lesssim 1$ regime due to the quite small superconducting gap achievable at the interface LaAlO\textsubscript{3}/SrTiO\textsubscript{3}. Indeed, in the split gate SQPC geometry implemented in Ref.~\cite{thierschmann2018ncomm}, a gap $\Delta\approx 22\mathrm{\mu}\text{eV}$ was observed, corresponding to a frequency $\nu\approx 5.3$ GHz, which perfectly fits the typical range where measurements in the electron quantum optics domain have been performed~\cite{dubois2013levitonsNature}. For more conventional superconducting materials it is in principle possible to reduce the gap by applying a magnetic field.

The adiabatic limit, where the energy scale related to the frequency of the drive is much smaller than the superconducting gap, is characterized by a very simple expression for the supercurrent, exhibiting Shapiro spikes whose height is proportional to the photoassisted amplitude of the drive considered. All other contributions to current and noise are strongly suppressed and ultimately vanish in this regime, since they involve quasiparticle transfers across the gap.

In conlcusion, our results extend the concept of Levitons as excitations minimizing the excess noise also in a superconducting background. Their peculiar features are best observed if the system is probed at frequencies bigger or at least comparable to the superconducting gap, {a condition which is within reach in nowadays experiments}. In the opposite regime, transport properties are dominated by {conventional} Shapiro spikes in the supercurrent, with a simple Josephson-like relation for any drive.

\begin{acknowledgements}
	M.A.\ and D.F.\ would like to thank N. Traverso Ziani for fruitful discussions.
	This work was granted access to the HPC resources of Aix-Marseille Universit\'e financed by the	project Equip@Meso (Grant No.\ ANR-10-EQPX29-01). It has been carried out in the framework of project ``one shot reloaded'' (Grant No.\ ANR-14-CE32-0017) and benefited from the support of the Labex ARCHIMEDE (Grant No.\ ANR11-LABX-0033) and the AMIDEX project (Grant No.\ ANR-11-IDEX-0001-02), funded by the ``investissements d'avenir'' French Government program managed by the French National Research Agency (ANR).
 \end{acknowledgements}
 
\appendix

\section{Nonequilibrium energy distributions}\label{app:distributions}
In this Appendix we connect the excess noise defined in Eq.~\eqref{eq:excess-noise-def} {of the main text} to the out-of-equilibrium energy distribution of electrons, commonly used in the context of electron quantum optics~\cite{bocquillon14}. Let us start by writing the model Hamiltonian with the explicit coupling to the external drive:
\begin{equation}
H=H_L+H_R+\lambda\sum_{\sigma=\up\down}(c_{L\sigma}^\dagger c_{R\sigma}+\text{H.c.})+eV(t)N_L\,.
\end{equation}
Here $c_{L/R\sigma}$ is the annihilation operator for the left/right lead at the point $x=0$ where the tunneling occurs and $N_L$ is the number operator for electrons in the left lead, the one where the voltage is applied. For our calculations it was convenient to include the effect of $V(t)$ into the tunneling amplitudes, as in Eq.\ \eqref{eq:H}. In order to do this it is sufficient to apply a unitary transformation generated by the operator
\begin{equation}
U=e^{ieN_L\int_{0}^{t}dt'\,V(t')}\,.
\end{equation}
Then the Hamiltonian transforms according to the relation $H\to UHU^\dagger+i\dot{U}U^\dagger$ and becomes
\begin{equation}
H=H_L+H_R+\lambda\sum_{\sigma=\up\down}\left[e^{i\varphi(t)}c_{L\sigma}^\dagger c_{R\sigma}+\text{H.c.}\right]\,,
\end{equation}
with $\varphi(t)=e\int_{0}^{t}dt'\,V(t')$\,. By including also the bare superconducting phase difference $\phi_0$ we finally obtain Eq.\ \eqref{eq:H}. Under the above transformation, electron operators of the left lead become
\begin{equation}
\tilde{c}_{L\sigma}=Uc_{L\sigma}U^\dagger=e^{-i\varphi(t)}c_{L\sigma}\,,
\label{eq:op-transf}
\end{equation}
while $c_{R\sigma}$ is unaffected. This shows that the effect of the external bias on the left lead electron operators {can be} encoded in the phase $\varphi(t)$.

We are now in position to compute nonequilibrium energy distributions of $L$-electrons and show how they connect with the excess noise. In the following we consider the effects of the drive $V(t)$ on the \emph{isolated} left electrode (meaning that we do not consider the coupling to the right one {consistently with the lowest order perturbation expansion discussed in the main text}). The building blocks of the calculation are the electron and hole coherence functions, which are the fundamental ingredients in electron quantum optics~\cite{bocquillon14}. They are defined as~\cite{grenier11,ferraro13} (since there is no dependence on the spin, the index $\sigma$ will be dropped in the following)
\begin{subequations}
	\begin{equation}
	\tilde{\mathcal{G}}^{(e)}(t,t')=\Braket{\tilde{c}_{L}^\dagger(x,t')\tilde{c}_{L}(x,t)}\,,
	\end{equation}
	\begin{equation}
	\tilde{\mathcal{G}}^{(h)}(t,t')=\Braket{\tilde{c}_{L}(x,t')\tilde{c}_{L}^\dagger(x,t)}\,,
	\end{equation}
	\label{eq:coherences}
\end{subequations}
where $x$ is any fixed position in the left electrode, where $V(t)$ is applied. Notice that the definition involves $\tilde{c}_L$ operators, since we want to describe nonequilibrium effects due to $V(t)$. By using Eq.\ \eqref{eq:op-transf}, coherence functions are expressed as
\begin{equation}
\tilde{\mathcal{G}}^{(e/h)}(t,t')=e^{\pm i[\varphi(t)-\varphi(t')]}\mathcal{G}^{(e/h)}(t,t')\,,
\end{equation}
where
\begin{equation}
\begin{split}
\mathcal{G}^{(e/h)}(t,t')&=-i\,g_0^{+-}(t-t')=+i\,g_0^{-+}(t'-t)\\
&=\int\frac{y\,dy}{w\sqrt{y^2-\Delta^2}}\Theta(y-\Delta)\,e^{iy(t-t')}
\end{split}
\end{equation}
are {zero temperature superconducting coherence functions at equilibrium, with no applied drive}. Notice that the {conventional free-fermion relation $\mathcal{G}^{(e)}(\tau)+\mathcal{G}^{(h)}(-\tau)\propto\delta(\tau)$ is recovered in the limit $\Delta\to 0$ as expected}.
Starting from Eq.\ \eqref{eq:coherences}, one can define energy distribution functions~\cite{ferraro13}
\begin{equation}
f^{(e/h)}(\omega)=\int_{-\mathcal{T}/2}^{\mathcal{T}/2}\frac{d\bar{t}}{\mathcal{T}}\int d\tau\,e^{i\omega\tau}\tilde{\mathcal{G}}^{(e/h)}\left(\bar{t}+\frac{\tau}{2},\bar{t}-\frac{\tau}{2}\right)\,,
\end{equation}
where $\mathcal{T}$ is the period of the drive. These quantities can be straightforwardly evaluated in terms of photoassisted coefficients. In particular, the equilibrium electron energy distribution is directly given by
\begin{equation}
f_\text{eq}(\omega)=-ig_0^{+-}(\omega)=2\pi\rho_0(\omega)n_\text{F}(\omega)=\frac{-2\omega\Theta(-\omega-\Delta)}{w\sqrt{\omega^2-\Delta^2}}\,,
\end{equation}
with $\rho_0(\omega)$ {properly defined} in Eq.\ \eqref{eq:rho-0} and the last expression being true at zero temperature. Finally, the complete hole energy distribution is found to be
\begin{equation}
\begin{split}
f^{(h)}(\omega)&=i\sum_{\ell\in\mathbb{Z}}|p_\ell|^2g_0^{-+}(-\omega-\Omega_\ell)\\
&=2\pi\sum_{\ell\in\mathbb{Z}}|p_\ell|^2\rho_0(-\omega-\Omega_\ell)n_\text{F}(\Omega_\ell-\omega)\\
&=\frac{2}{w}\sum_{\ell\in\mathbb{Z}}|p_\ell|^2\frac{(-\omega-\Omega_\ell)\Theta(-\omega-\Omega_\ell-\Delta)}{\sqrt{(-\omega-\Omega_\ell)^2-\Delta^2}}\,.
\end{split}
\end{equation}
Thus the connection in Eq.\ \eqref{eq:excess-noise-f} of the main text is established.

\section{General expressions for current and noise at low temperature}\label{app:is}
In this Appendix we give general expressions for dc current and noise at low but finite temperature, in terms of a single integral over energy. Before that, let us briefly recall how full Green's functions $\hat G$ are related to unperturbed ones $\hat g$ via Dyson's equations. The simplest equation is the one for the advanced and retarded Green's functions and reads:
\begin{equation}
\hat{G}^{a/r}(t,t')=\hat{g}^{a/r}(t-t')+\int d\tau\hat{g}^{a/r}(t-\tau)\hat\Sigma^{a/r}(\tau)\hat{G}^{a/r}(\tau,t')
\label{eq:dys1}
\end{equation}
where $\hat\Sigma^{a/r}$ are the self-energy matrices. In our case, they are simply $\hat{\Sigma}^{a/r}_{LL}=\hat{\Sigma}^{a/r}_{RR}=0$ and $=\hat{\Sigma}^{a/r}_{LR}=[\hat{\Sigma}^{a/r}_{RL}]^\dagger=\hat{\mathcal{W}}$, with the matrix $\hat{\mathcal{W}}$ given in \eqref{eq:w}. The equation for $\hat{G}^{+-}$ is more complicated:
\begin{equation}
\hat{G}^{+-}=\hat{g}^{+-}+\hat{G}^r\hat{\Sigma}^r\hat{g}^{+-}+\hat{G}^{+-}\hat{\Sigma}^a\hat{g}^a\,,
\end{equation}
where a convolution over intermediate time arguments is assumed, like in Eq.\ \eqref{eq:dys1}. From this expression we obtain
\begin{subequations}
	\begin{align}
	\hat{G}_{LR}^{+-}&=\hat{g}^{+-}\hat{\mathcal{W}}\hat{G}^{a}_{RR}+\hat{g}^r\hat{\mathcal{W}}\hat{G}^{+-}_{RR}\,,\\
	\hat{G}_{RL}^{+-}&=\hat{G}_{RR}^r\hat{\mathcal{W}}^\dagger\hat{g}^{+-}+\hat{G}_{RR}^{+-}\hat{\mathcal{W}}^\dagger\hat{g}^{a}\,,
	\end{align}
	\label{eq:dys2}
\end{subequations}
where a convolution is again implied.
We can now use these relations into Eqs.~\eqref{eq:i-start}--\eqref{eq:s-start} {in the main text} and truncate the expansion at lowest order in $\lambda$ to obtain the following general expressions:
	\begin{subequations}
		\begin{equation}
		\begin{aligned}
		I_0&=4\pi e\lambda^2\sum_{\ell\in\mathbb{Z}}|p_{\ell}|^2\int d\omega\rho_0(\omega)\rho_0(\omega-\Omega_{\ell})\\
		&\quad\times[n_\text{F}(\omega-\Omega_{\ell})-n_\text{F}(\omega)]\,,
		\end{aligned}
		\end{equation}
		\begin{equation}
		\begin{aligned}
		I_1&=4\pi e\lambda^2\sum_{\ell\in\mathbb{Z}}\text{Re}\left[e^{i\phi_0}p_\ell p_{-\ell-2q}\right]\int d\omega\rho_1(\omega)\rho_1(\omega-\Omega_{\ell})\\
		&\quad\times[n_\text{F}(\omega-\Omega_{\ell})-n_\text{F}(\omega)]\,,
		\end{aligned}
		\end{equation}
		\begin{equation}
		\begin{aligned}
		I_J&=-4\pi e\lambda^2\sum_{\ell\in\mathbb{Z}}\text{Im}\left[e^{i\phi_0}p_\ell p_{-\ell-2q}\right]\int d\omega\rho_1(\omega)n_\text{F}(\omega)\\
		&\quad\times[\rho_2(\omega+\Omega_\ell)+\rho_2(\omega-\Omega_\ell)]\,,
		\end{aligned}
		\end{equation}
		\begin{equation}
		\begin{aligned}
		S_0&=8\pi e^2\lambda^2\sum_{\ell\in\mathbb{Z}}|p_\ell|^2\int d\omega\rho_0(\omega)\rho_0(\omega-\Omega_\ell)\\
		&\quad\times[n_\text{F}(\omega)n_\text{F}(\Omega_\ell-\omega)+n_\text{F}(-\omega)n_\text{F}(\omega-\Omega_\ell)]\,,
		\end{aligned}
		\end{equation}
		\begin{equation}
		\begin{aligned}
		S_1&=8\pi e^2\lambda^2\sum_{\ell\in\mathbb{Z}}\text{Re}\left[e^{i\phi_0}p_\ell p_{-\ell-2q}\right]\int d\omega\rho_1(\omega)\rho_1(\omega-\Omega_\ell)\\
		&\quad\times[n_\text{F}(\omega)n_\text{F}(\Omega_\ell-\omega)+n_\text{F}(-\omega)n_\text{F}(\omega-\Omega_\ell)]\,.
		\end{aligned}
		\end{equation}
	\label{eq:is-general}
	\end{subequations}
Functions appearing in the above integrals are defined in terms of the unperturbed Green's functions $\hat{g}^{a/r}(\omega)=g^{a/r}_0(\omega)\hat{\sigma}_0+g_1^{a/r}(\omega)\hat{\sigma}_1$ given in Eq.~\eqref{eq:greens} and are
\begin{equation}
\begin{split}
\rho_0(\omega)&=\frac{1}{\pi}\text{Im}[g_0^a(\omega)]=\frac{|\omega|}{\pi w\sqrt{\omega^2-\Delta^2}}\Theta(|\omega|-\Delta)\,,\label{eq:rho-0}\\
\rho_1(\omega)&=\frac{1}{\pi}\text{Im}[g_1^a(\omega)]=\frac{-\Delta\text{sgn}(\omega)}{\pi w\sqrt{\omega^2-\Delta^2}}\Theta(|\omega|-\Delta)\,,\\
\rho_2(\omega)&=\frac{1}{\pi}\text{Re}[g_1^a(\omega)]=\frac{\Delta}{\pi w\sqrt{\Delta^2-\omega^2}}\Theta(\Delta-|\omega|)\,.
\end{split}
\end{equation}
Notice that in the above results, the dependence on temperature is confined to Fermi functions. This is because we assume that the temperature is low enough for the gap $\Delta$ to be considered constant.
The evaluation of integrals in Eq.\ \eqref{eq:is-general} at zero temperature yields the results presented in the main text, see equations Eqs.~\eqref{eq:i0}--\eqref{eq:s1}.

\end{document}